\documentclass[conference]{IEEEtran}

\usepackage{cite}
\usepackage{amsmath,amssymb,amsfonts}
\usepackage{algorithmic}
\usepackage{graphicx}
\usepackage{textcomp}
\usepackage{xcolor}
\usepackage{caption}

\begin{document}

\title{Terraform --- Automating Infrastructure As A Service\\
}

\author{\IEEEauthorblockN{Michael Howard}
\IEEEauthorblockA{\textit{Computer Science Department} \\
\textit{Portland State University}\\
Portland, USA \\
mihoward@pdx.edu}
}

\maketitle

\begin{abstract} \label{sec:abs}
Developing a software service requires a strict software development life cycle and process.  This process demands controlling all application code through source control management as well as a rigorous versioning and branching strategy.  However, the platform and infrastructure also benefit from this rigor.  Software services must be deployed to a target run time environment and provisioning that environment through manual user actions is tedious and error-prone.  Provisioning manually also becomes prohibitive as the number of resources grow and spread globally over multiple regions.  The answer is to apply the same rigor to provisioning the infrastructure as applied to developing the application software.  Terraform provides a platform allowing infrastructure resources to be defined in code.  This code not only allows the automation of the infrastructure provisioning but also allows for a strict development and review life cycle, same as the application software.    
\end{abstract}


\section{Introduction} \label{sec:intro}
Terraform is an Infrastructure As Code (IaC) client tool developed by HashiCorp.  It allows the user to define both cloud and on-premise compute resources in human-readable configuration files.  These files are created using the HashiCorp Configuration Language (HCL).  The syntax is declarative with each block of code defining a resource to be provisioned.  Declarative definitions (versus imperative) allow the user to define the desired state, rather than an exhaustive list of all the interim steps required to achieve that state.

Section \ref{sec:workflows} discusses typical Terraform workflows and how the tool is used.  Section \ref{sec:language} digs more into the structure of the HCL code.  Various platforms that can run the Terraform code are discussed in Section \ref{sec:env} while the provider plugins are expanded upon in Section \ref{sec:providers}.  Section \ref{sec:cdk} details the Cloud Development Kit interfacing Terraform to multiple high-level programming languages.  The alternatives to Terraform are examined in Section \ref{sec:alternatives}.  Other related topics of research are discussed in \ref{sec:related} which leads to a summary discussion in Section \ref{sec:conclusion}.

\section{Basic Workflows} \label{sec:workflows}
\begin{figure}[htbp]
    \centerline{\includegraphics[width=.9\linewidth, keepaspectratio]{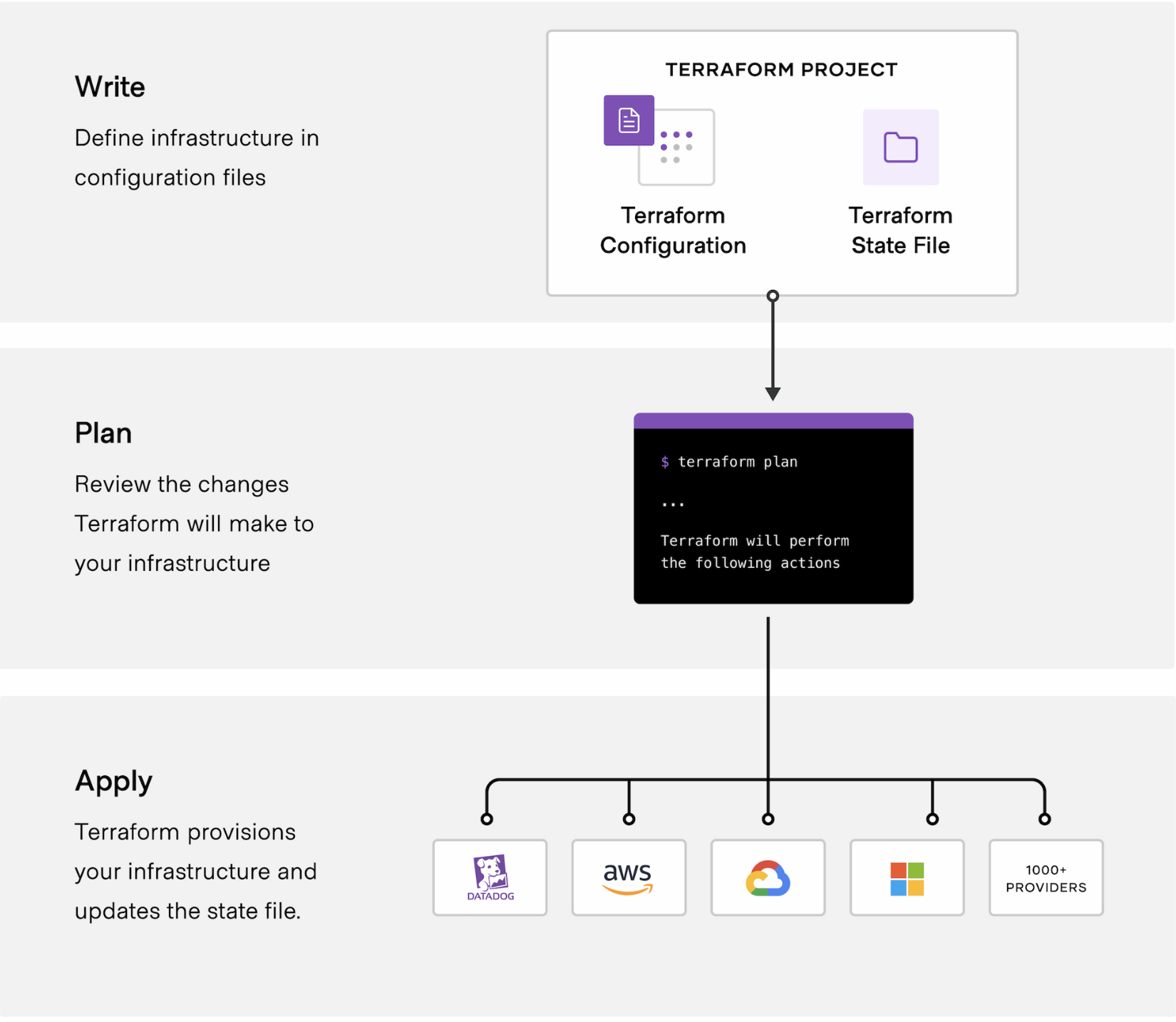}}
    \captionsetup{width=.8\linewidth}
    \caption{High-level workflow for Terraform covering write, plan and apply stages.}
    \label{fig:workflow}
\end{figure}

Figure \ref{fig:workflow} introduces the basic workflow for Terraform.  The development and implementation of IaC is broken into write, plan and apply stages.

\subsection{Write}
The \emph{write} stage focuses on developing the code required to drive the plan.  The infrastructure resources are defined here for the providers and services required.  Multiple providers may be included.  The code is contained in configuration files with a .tf extension.  There is an additional option to store the configuration in JavaScript Object Notation (JSON) format which then requires the .tf.json extension.  The correct extension must be used in order for the Terraform tool to detect the configuration file while generating the plan.

For example, a user wishes to create a mini-cluster of virtual machines in Amazon Web Service (AWS).  Three virtual machines, a Virtual Private Cloud (VPC) network, security group and a load balancer service are required for this cluster.  The Terraform configuration code declares each of these resources separately with corresponding parameters.  Section \ref{sec:language} goes into more detail of the configuration language.

\subsection{Plan}
Once the configuration code has been written, the next stage is to run the Terraform tool to generate a \emph{plan}.  The tool is run either through the local Command Line Interface (CLI) or via other high-level language that can interface to the Terraform framework via the Cloud Development Kit (CDK).  Section \ref{sec:cdk} digs into further detail of the CDK.  Running the Terraform plan will scan local directories for configuration files ending in .tf or .tf.json and process these into a list of actions to be sent to the provider(s).  This list is called the execution plan and encompasses all create, update and destroy actions needed to make the target infrastructure match what is declared in the configuration code.

The plan generation additionally has a dependency on the existing infrastructure that is represented in the \emph{state}.  The state details all infrastructure resources that are currently present.  The state file exists either locally in the file system or remotely.  For example, the configuration contains an Elastic Compute (EC2) virtual machine named ``VirtualMachine1''.  Upon running Terraform to generate the plan, the current state file is checked.  If the VirtualMachine1 EC2 is already existing, the plan does not create it.  It will either be an update action or no operation.

\subsection{Apply}
\begin{figure}[htbp]
    \centerline{\includegraphics[width=.9\linewidth, keepaspectratio]{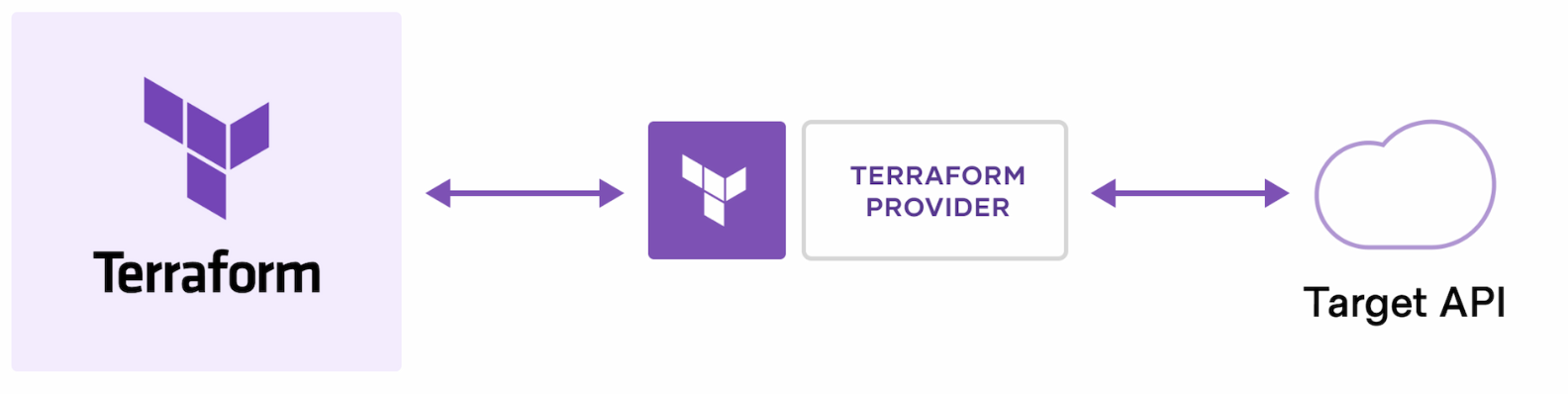}}
    \captionsetup{width=.8\linewidth}
    \caption{Terraform actions go through a provider module to translate into API calls specific to that provider.}
    \label{fig:target}
\end{figure}

The final stage in the Terraform workflow is to \emph{apply}.  Running the tool on a plan executes each action against the corresponding provider.  Figure \ref{fig:target} shows Terraform interacting with its provider plugin which subsequently calls into the Application Programming Interface (API) of the corresponding cloud provider (e.g. AWS).  Provider modules act as the abstraction between the configuration code and unique API defined by each infrastructure provider.  The providers are further discussed in Section \ref{sec:providers}.  As part of apply, the state is updated to represent the changes in the target infrastructure.  

\section{Configuration Language} \label{sec:language}
Terraform scans configuration files and generates a corresponding plan.  The configuration files are written in the HashiCorp Configuration Language (HCL).  This language is declarative.  Declarative offers an advantage over imperative in that the desired state of the infrastructure can be directly coded.  An imperative language requires defining all the interim steps to arrive at the desired state.

\begin{figure}[htbp]
    \centerline{\includegraphics[width=.9\linewidth, keepaspectratio]{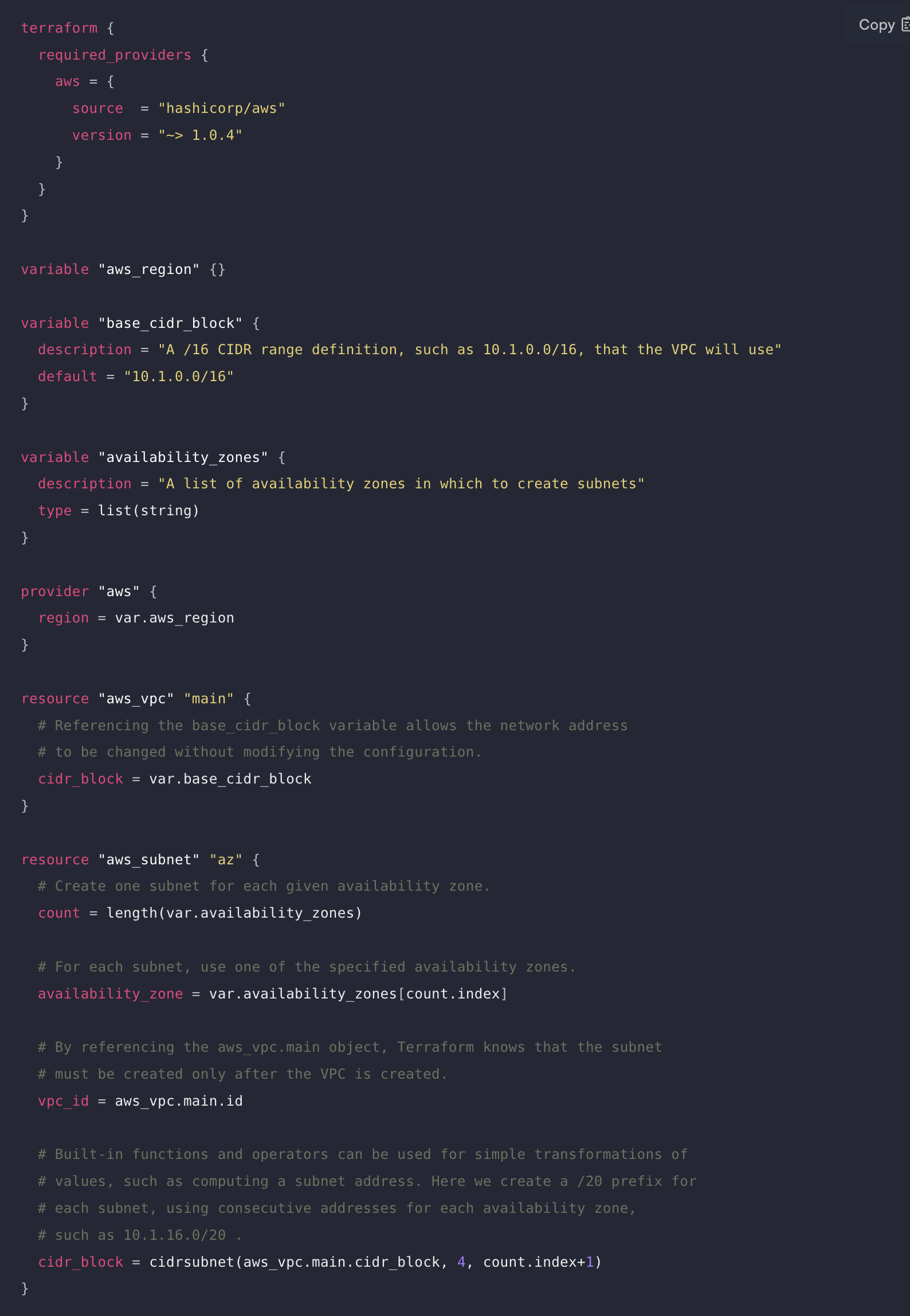}}
    \captionsetup{width=.8\linewidth}
    \caption{An HCL example declaring the required provider, a VPC and subnet.}
    \label{fig:hcl}
\end{figure}

The main purpose of HCL is to define resources.  The code is written in blocks with each block representing an infrastructure object.  A Terraform configuration is a complete document in HCL telling Terraform how to manage a given collection of infrastructure resources.  Figure \ref{fig:hcl} shows example code that declares the required provider plugin as well as a VPC and corresponding subnet.  The purpose of the block is defined by the block type.  A \emph{variable} block type is used as a parameter in other blocks.  A \emph{provider} defines what provider plugin is used to translate the resource block into API calls to the infrastructure provider.  The \emph{resource} block is used to define a concrete resource in the provider's infrastructure.  Resource blocks are translated into create, update or delete API calls to the provider's target infrastructure service.

\section{Framework Environment} \label{sec:env}
Environments that can run Terraform are the CLI, Terraform Cloud, Terraform Enterprise and CDK.  The CLI is the most common.  Pre-built binaries can be downloaded or the Golang source code\footnote{https://github.com/hashicorp/terraform} can be cloned and built.  The Terraform tool runs on a local set of configuration files.  These files can be organized into subdirectories which Terraform will automatically traverse.  The \emph{state} file is typically generated in the same directory that the tool runs from.  However, there is a \emph{remote} option which generates state files in a remote, central location such that multiple Terraform clients may apply their plans and still synchronize their view of the existing infrastructure.  

\emph{Terraform Cloud} is a Terraform environment hosted by HashiCorp.  As a hosted service, users log in to generate and apply plans.  The Terraform tool-chain itself is maintained by HashiCorp while the state files are centrally stored such that all users are running against the same current infrastructure state.  \emph{Terraform Enterprise} is a self-hosted version of Terraform Cloud.  If offers the same cloud-based feature set but is designed to be deployed within an enterprise's private cloud.

Another version of the CLI or local environment is the \emph{Cloud Development Kit (CDK)}.  Rather than running the CLI tool directly, CDK permits five supported high-level languages to generate and apply Terraform plans.  Code in these supported languages is able to call in to the Terraform framework, replacing the Terraform CLI.  Section \ref{sec:cdk} provides further details on CDK.

\section{Providers} \label{sec:providers}
Terraform relies on plugins called \emph{providers} to interact and abstract the various infrastructure providers.  Each provider must be declared in the configuration using the ``provider'' block type.  Once a provider has been declared, the corresponding plugin is included while generating the plan.  Declared resources utilize the provider's underlying API to perform create, update and delete actions needed to ensure the resource ends up in the desired state. 

Providers come from a publicly available registry of known plugins\footnote{https://registry.terraform.io/browse/providers}.  The list is extensive and covers all known cloud, Software as a Service (SaaS) and other APIs.  These provider plugins allow the resource and data source blocks to be declared without needing details on the specific provider's API.  Each provider maintains documentation on the Terraform blocks it supports along with the corresponding parameters.

\section{Cloud Development Kit} \label{sec:cdk}
The Cloud Development Kit (CDK) for Terraform allows the use of other programming language to define and provision infrastructure.  CDK gives access to the entire Terraform ecosystem without requiring development in HashiCorp Configuration Language (HCL) and running it via the CLI tool.  Additionally, a user can more easily integrate with an existing tool-chain for testing and dependency management.  The following languages are currently supported:
\begin{itemize}
    \item Typescript
    \item Python
    \item Java
    \item C\#
    \item Go
\end{itemize}

\begin{figure}[htbp]
    \centerline{\includegraphics[width=.9\linewidth, keepaspectratio]{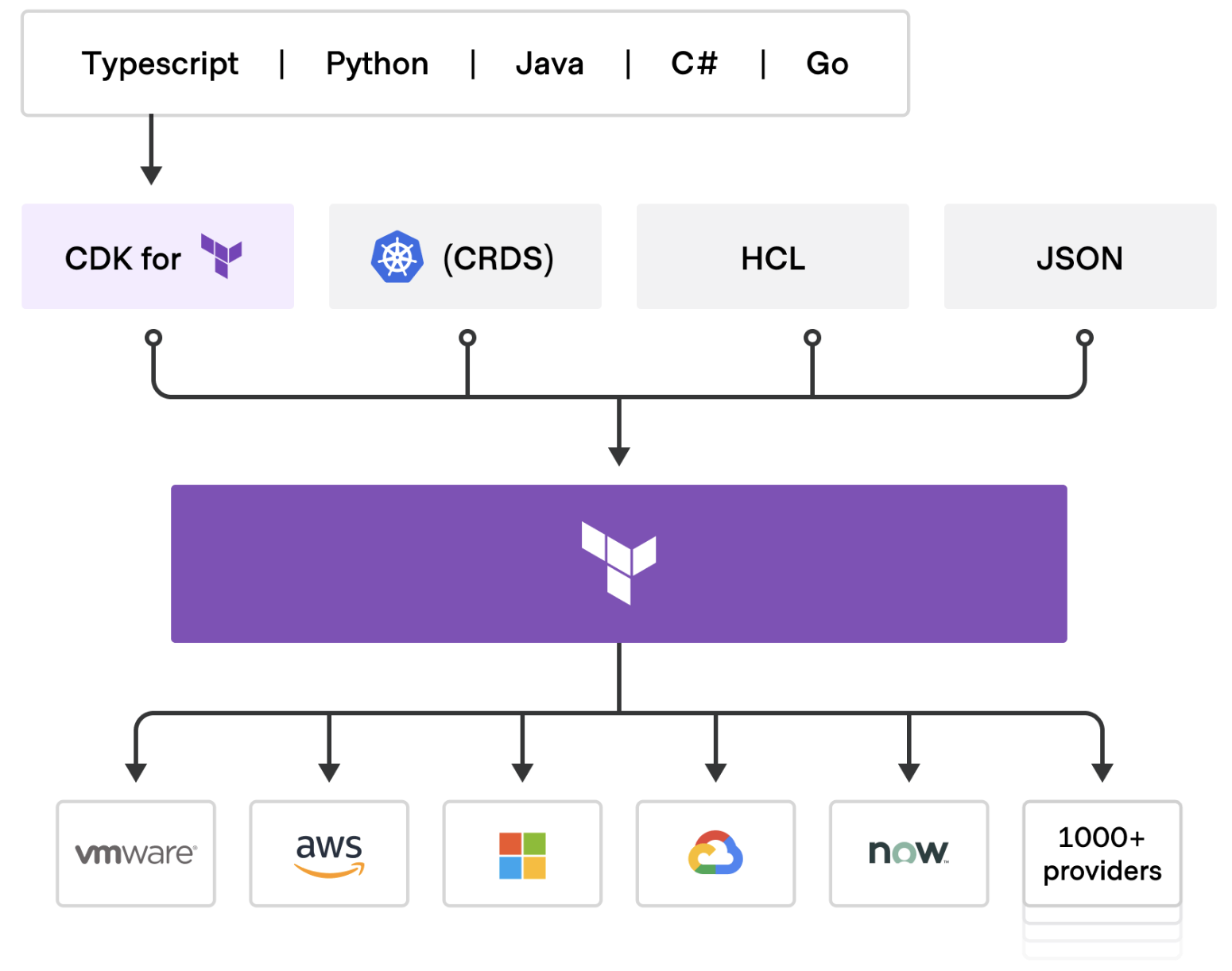}}
    \captionsetup{width=.8\linewidth}
    \caption{CDK and other pathways to define configuration, input to Terraform and provision infrastructure through multiple providers.  Configuration input may be through CDK, CRDS, HCL or JSON.}
    \label{fig:cdk}
\end{figure}

Figure \ref{fig:cdk} shows the various input pathways to Terraform.  CDK may be invoked from its five supported languages while configuration code in HCL or JSON require the Terraform CLI.  Kubernetes' Custom Resource Definitions (CRDS) are another possibility but will not be covered here.

\section{Alternatives} \label{sec:alternatives}
Terraform provides an abstraction of providers and their resources.  It can represent physical hardware, virtual machines, containers, network configurations, email and Domain Name Service (DNS) providers.  Given the breadth of resources and providers, Terraform does overlap with other tools.  Some of these tools will be discussed here.

\subsection{Chef and Puppet}
Chef\footnote{https://www.chef.io/} and Puppet\footnote{https://puppet.com/} are configuration management tools.  They are designed to install and manage software on compute resources that already exist.  Terraform instead focuses on the bootstrapping and initializing of those compute resources.  It works well in conjunction with configuration management.

\subsection{CloudFormation and Heat}
CloudFormation\footnote{https://aws.amazon.com/cloudformation/} and Heat\footnote{https://docs.openstack.org/heat/latest/} are both tools that represent infrastructure as code, just like Terraform.  The configuration files allow the infrastructure to be elastically created, modified and destroyed.  The big advantage which Terraform provides is it is provider-agnostic.  CloudFormation is an Amazon Web Service (AWS) tool and only works with provisioning other AWS resources.  Heat similarly operates only on an OpenStack API.  Terraform not only supports multiple providers but can also combine resources from each into a single plan.  Thus, it introduces \emph{multi-cloud} provisioning.

Another feature which Terraform has over CloudFormation and Heat is the separation of the plan and execution.  Terraform has the distinct stage to generate a plan which also takes into account the existing state of the infrastructure.  The plan is then optionally reviewed and approved before the apply stage executes each plan action.  Terraform also has a graph feature which displays the plan actions ordered by dependency.    

\subsection{Boto and Fog}
Boto\footnote{http://boto.cloudhackers.com/en/latest/} and Fog\footnote{https://github.com/fog/} are similar to provider plugins in Terraform.  They abstract the API to a particular infrastructure provider.  Both still require some high level programming language to declare the resources and call in to their respective libraries.  In contrast, Terraform abstracts this functionality as an extensive set of plugins and providers the high-level configuration language to allow declaring the resources in a provider-agnostic manner.

\section{Related Work} \label{sec:related}
The Organization for the Advancement of Structured Information Standards (OASIS)\footnote{https://www.oasis-open.org/} was founded in 1993 as a non-profit consortium that works on the development, convergence and adoption of open standards for cybersecurity, cloud computing and related areas\cite{oasis}.  The two standards that are relevant to this paper are Topology and Orchestration Specification for Cloud Applications (TOSCA) and Cloud Application Management for Platforms (CAMP).

\subsection{TOSCA}
TOSCA is an open standard that describes a topology of cloud-based web services, their components, relationships and the processes that manage them.  Version 1.0 was approved 16 January 2014 by OASIS.  The standard enables portability and automated management across cloud providers regardless of the underlying infrastructure.  This standard improves reliability and reduces cost while facilitating the continuous delivery of applications across their entire lifecycle.    

\begin{figure}[htbp]
    \centerline{\includegraphics[width=.9\linewidth, keepaspectratio]{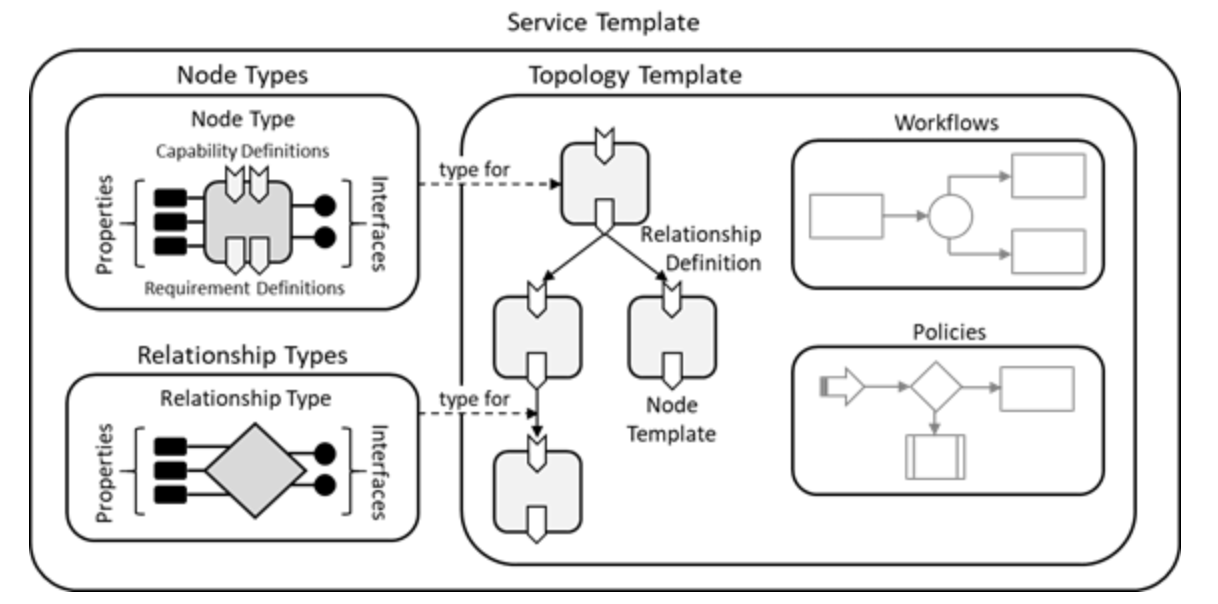}}
    \captionsetup{width=.8\linewidth}
    \caption{Structural Elements of a TOSCA Service Template and its Relations.}
    \label{fig:tosca}
\end{figure}

Version 2.0 is the current and was approved 28 October 2020 \cite{tosca_v2}.  The core specification provides a language for describing service components and their relationships using a service topology, and it provides for specifying the lifecycle management procedures that allow for creation or modification of services using orchestration processes.  A TOSCA \emph{Service Template}, as shown in Figure \ref{fig:tosca}, combines topology and orchestration needed in different environments to enable automated deployment of services and their management throughout the complete service lifecycle.

The TOSCA language has the ability to automate lifecycle management for the following:
\begin{itemize}
    \item Infrastructure as a Service deployments for multiple cloud providers (i.e. OpenStack, AWS, Microsoft Azure).
    \item Deploy containerized applications to existing orchestrators (i.e. Kubernetes).
    \item Define the management of Virtual Network Functions.
    \item Support on-demand creation of network services.
    \item Define Functions as a Service applications that can run without any corresponding deployment.
    \item Deploy services to Internet of Things (IoT) and Edge devices while minimizing latency.
    \item Support open and interoperable process control architectures.
\end{itemize}

Implementations of the TOSCA standard can take the form of:
\begin{itemize}
    \item Source code: a Yet-Another-Markup-Language (YAML) document which defines the Service Template.
    \item Processor: a tool or engine to parse the Service Template document.
    \item Orchestrator: a tool or engine that processes the Service Template in order to deploy and manage an application.
    \item Translator: a tool to translate the Service Template into another language such as Helm Charts or Amazon CloudFormation templates.
    \item Generator: a tool to generate a Service Template.
    \item Archive: Cloud Service Archive (CSAR), a package containing the Service Template and other artifacts needed for deployment.  
\end{itemize}

Links to all known TOSCA implementations are listed in the oasis-open repository in GitHub\footnote{https://github.com/oasis-open/tosca-community-contributions/wiki/Known-TOSCA-Implementations}.  Also, a full list of TOSCA technical committee members may be viewed at the oasis-open membership page\footnote{https://www.oasis-open.org/committees/membership.php?wg\_abbrev=tosca}.

\subsection{CAMP}
The Cloud Application Management for Platforms (CAMP) is another standard to come from the OASIS consortium. Its technical committee published version 1.0 in August 2012 and was a collaboration between CloudBees, Cloudsoft Corporation, Huawei, Oracle, Rackspace, Red Hat, and Software AG \cite{camp}.  The technical committee was closed by OASIS on 23 April 2021 and is no longer active.  As CAMP is referenced frequently in cloud computing research, we will include it here for completeness.

The CAMP technical committee's goal advances an interoperable protocol that packages and deploys cloud-hosted applications.  The standard defines models, mechanisms and protocols for the management of a Platform as a Service (PaaS) environment.  PaaS describes a service where the users manage the platform that applications are hosted on.  In contrast, the TOSCA standard focuses on Infrastructure as a Service (IaaS).  PaaS exists a level above IaaS and thus CAMP and TOSCA are complementary standards rather than overlapping.

\section{Conclusion and Discussion} \label{sec:conclusion}
In this paper, we discuss Terraform as an Infrastructure as Code (IaC) tool.  Its framework allows the user to declare infrastructure resources through code, generate an execution plan from that code and finally apply the plan.  Applying works through provider plugin modules which translate the execution actions into API calls specific to the provider.  Not only is the infrastructure created in a programmatic manner, it is agnostic of the underlying provider (i.e. AWS, private cloud, VMWare, Microsoft Azure).  This abstraction avoids vendor lock-in and increases portability between vendors.

The Terraform configuration language is discussed as well as the possible environments that can interpret and run the corresponding code.  These environments range from local execution to both cloud and private hosted services.  Provider modules allow interfacing to almost any third-party infrastructure vendor or service while the Cloud Development Kit (CDK) enables a user to integrate with the Terraform framework from five popular high-level programming languages.  Configuration management tools such as Chef, Ansible and Puppet focus more on automating the software and configurations within an infrastructure resource and thus, do not directly compete with Terraform.  Amazon's CloudFormation and OpenStack's Heat do compete.  However, their framework only supports their own infrastructure and cannot provision resources from other providers.

As of this writing, Terraform focuses on the provisioning of infrastructure and services.  The building and deployment of software applications to that infrastructure requires additional tool-chains and automation.  A product that overseas an entire operation from multi-cloud infrastructure provisioning, application deployment and runtime orchestration does not exist at this time.  One promising possibility is Topology and Orchestration Specification for Cloud Applications (TOSCA).  

TOSCA is a standard developed by the Organization for the Advancement of Structured Information Standards (OASIS) consortium.  Part of the standard is a specification language that allows users to create YAML-based Service Templates.  These templates declare nodes, workflows and relationships such that a more complete picture may be specified for the \emph{total} operation of an application and its required infrastructure.  TOSCA implementations include template code interpreters, orchestrators, translators and archive tools.  OpenStack Heat is an example of a TOSCA-compatible framework but is still limited in its breadth and focuses mostly on the provisioning.  Further research in TOSCA is required, both to expand and mature the standard as well as bring a new suite of framework tools into the mainstream that implement it.

\vspace{12pt}

\end{document}